\documentclass[twocolumn,showpacs,preprintnumbers,amsmath,superscriptaddress,amssymb,prb]{revtex4}

\usepackage{dcolumn}% Align table columns on decimal point
\usepackage{upgreek} %\upmu
\usepackage{txfonts} %\muup
\usepackage{ifpdf}
\ifpdf
  \usepackage[pdftex]{graphicx}
  \usepackage[pdftex, pdfpagelayout=TwoColumnRight, pdfstartview=FitH, plainpages=false, pdfpagelabels]{hyperref}
 \else
   \usepackage[dvips]{graphicx}

\fi % Used for figures
\usepackage{enumerate}

\begin{document}

\title{Observation of superluminal geometrical resonances in Bi$_2$Sr$_2$CaCu$_2$O$_{8+x}$ intrinsic Josephson junctions}

\author{S. O. Katterwe}
\author{A. Rydh}
\author{H. Motzkau}
\affiliation{Department of Physics, Stockholm University, AlbaNova
University Center, SE-10691 Stockholm, Sweden}

\author{A. B. Kulakov}

\affiliation{ Institute of Solid State Physics, Russian Academy of
Sciences, 142432 Chernogolovka, Russia}

\author{V. M. Krasnov}
\email{Vladimir.Krasnov@fysik.su.se}
\affiliation{Department of
Physics, Stockholm University, AlbaNova University Center,
SE-10691 Stockholm, Sweden}

\date{\today}

\begin{abstract}
We study Fiske steps in small $\mathrm{Bi_2Sr_2CaCu_2O}_{8+x}$
mesa structures, containing only few stacked intrinsic Josephson
junctions. Careful alignment of magnetic field prevents
penetration of Abrikosov vortices and facilitates observation of a
large variety of high quality geometrical resonances, including
superluminal with velocities larger than the slowest velocity of
electromagnetic waves. A small number of junctions limits the
number of resonant modes and allows accurate identification of
modes and velocities. It is shown that superluminal geometrical
resonances can be excited by subluminal fluxon motion and that
flux-flow itself becomes superluminal at high magnetic fields. We
argue that observation of high-quality superluminal geometrical
resonances is crucial for realization of the coherent flux-flow
oscillator in the THz frequency range.

\end{abstract}


\maketitle

\section{Introduction}
%**********************
Stacked Josephson junctions represent a macroscopic
electromagnetic system which can be easily tuned from Lorentz
invariant (uncoupled, single junctions \cite{Laub}), to a
non-invariant state by decreasing the layer thickness $d$ below
the magnetic screening length $\lambda_\mathrm{S}$. The lack of
Lorentz invariance is caused by the absence of local relation
between electric and magnetic fields. \cite{Fluxon} In thin-film
stacks, $d \ll \lambda_\mathrm{S}$, magnetic field is non-local
and is created cooperatively by the whole stack, leading to
inductive coupling of junctions.\cite{UstFiske,SakUst} The
strongest coupling is achieved in atomic scale intrinsic Josephson
junctions (IJJs), naturally formed in
$\mathrm{Bi_2Sr_2CaCu_2O}_{8+x}$ (Bi-2212) high-$T_\mathrm{c}$
superconductors. The lack of Lorentz invariance leads to a number
of unusual electrodynamic properties, such as splitting of the
dispersion relation of electromagnetic waves,\cite{KleinerModes}
and a possibility of superluminal (faster than the slowest
electromagnetic wave velocity) fluxon motion, accompanied by
Cherenkov-like radiation. \cite{Fluxon,Modes,Cherenkov,Savelev}
The coupling facilitates phase-locking of junctions, which may
lead to coherent amplification of the emission
power\cite{Barbara,SakUst,Shitov} $\propto N^2$, where $N$ is the
number of junctions. IJJs allow easy integration of many strongly
coupled stacked junctions. Furthermore, the large energy gap in
Bi-2212\cite{SecondOrder} facilitates operation in the important
THz frequency range. Therefore, IJJs are intensively studied as
possible candidates for realization of a coherent THz
oscillator.\cite{Savelev,Batov,Ozyuzer,Bae2007,FFlowSimul,FFlowMachida,FFlowKoshelev,FFlowRyndyk,
NonEquil,Bul,Hu,TTachiki,MTachiki,Klemm,TheoryFiske}

The flux-flow oscillator (FFO) is based on regular motion of
Josephson vortices (fluxons).\cite{Koshelets} To facilitate
coherent emission from a stacked FFO, fluxons in the stack must be
arranged in a rectangular (in-phase) lattice. It can be stabilized
by geometrical confinement in small Bi-2212 mesa
structures\cite{Katterwe} at high magnetic fields $B \sim \Phi_0/
s \lambda_\mathrm{J} \simeq 2\,\mathrm{T}$, where $\Phi_0$ is the
flux quantum, $s=1.55\,\mathrm{nm}$ is the interlayer spacing, and
$\lambda_\mathrm{J}$ is the Josephson penetration depth of IJJs.
At lower fields flux-flow is metastable or
chaotic.\cite{Modes,Compar} Yet, the rectangular lattice is
insufficient for achieving high emission power from the FFO. Since
the power is proportional to the square of the electric field
amplitude $E_\mathrm{ac}$, a large $E_\mathrm{ac}$ should also be
established.\cite{TheoryFiske} In a single junction
FFO\cite{Koshelets} this is achieved via Lorentz contraction of
fluxons at the velocity matching condition.\cite{Laub} However,
such a mechanism is not available in stacked
junctions\cite{FluxonDouble} due to the absence of Lorentz
invariance in the system.\cite{Fluxon} Therefore, powerful
emission is only possible in the presence of high quality $Q \gg
1$ geometrical (Fiske) resonances, which would amplify
$E_\mathrm{ac}\propto Q$.\cite{TheoryFiske} Thus, observation of
high-speed and high-quality geometrical resonances is a
prerequisite for realization of the high-power coherent FFO.

For artificial low-$T_\mathrm{c}$ stacked Josephson junctions both
velocity splitting\cite{UstFiske,SakUst} and coherent
amplification of radiation\cite{Shitov} were observed
experimentally. However, for IJJs only the slowest out-of-phase
Fiske steps have been unambiguously detected.
\cite{Fiske,FiskeSlow,FFLatyshev} Observation of superluminal
resonances in IJJs may be obscured by several factors: for stacks
with a large number of IJJs, $N \sim 100$, the velocity splitting
may be too dense to be identified and the exact number of
junctions in the flux-flow state difficult to estimate. The
necessity to operate in large magnetic fields may also lead to
intrusion of Abrikosov vortices that distort fluxon order and
dampen resonances.\cite{Katterwe,Bae2009} Coherently emitting
resonant modes can be dampen by radiative losses\cite{TheoryFiske}
and, as discussed below, high frequency resonances are very
sensitive to a spread in junction resistances.

In this work we study Fiske steps in small Bi-2212 mesa structures
containing only few IJJs. Careful alignment of magnetic field
parallel to CuO planes obviates intrusion of Abrikosov vortices
and leads to observation of a large variety of high quality
geometrical resonances with different velocities. Small $N$ limits
the number of resonant modes and simplifies their identification.
We demonstrate both experimentally and numerically that
superluminal geometrical resonances are excited by subluminal
flux-flow. Simultaneously we observe that flux-flow itself becomes
superluminal at high magnetic fields. Finally we observe an
asymmetry between even and odd resonance modes, which can be taken
as indirect evidence for considerable flux-flow emission from the
stack.\cite{TheoryFiske}

%The paper is organized as follows: In Sec.~\ref{sec:generalRelations} we provide explicit
%estimation of characteristic velocities of electromagnetic waves
%in IJJs and the range of resonant modes and Fiske step voltages,
%reachable in the flux-flow state. In Sec.~\ref{sec:experimental} we describe details
%of sample fabrication and measurements. In Sec.~\ref{sec:resultsDiscussion} we present our
%main experimental results and analyze the observed Fiske steps. In
%Sec.~\ref{sec:numericalModelling} we present numerical simulations which confirm the
%possibility of excitation of superluminal resonant modes by
%subluminal flux-flow. Finally, in Sec.~\ref{sec:qualityFactor} we discuss the
%significance of superluminal high-$Q$ geometrical resonance for
%achieving high emission power from the stacked FFO.

\section{General relations}
%**********************
\label{sec:generalRelations} Inductive coupling of $N$ stacked
Josephson junctions leads to splitting of the dispersion relation
of electromagnetic waves into $N$ branches with characteristic
velocities\cite{KleinerModes}
% *** EQUATION ***
\begin{equation}\label{Cmodes}
\frac{c_n}{c_0}=\left[1-2S\cos{\frac{n\pi}{N+1}}\right]^{-1/2},\qquad({n=1,2,\ldots,N}).
\end{equation}
% *** END EQUATION ***
Here $c_0$ is the (Swihart) velocity of light in a single junction
and $S\simeq 0.5/\cosh(d/\lambda_\mathrm{S})$ is the coupling constant.
Since there has been some confusion about values of extremal velocities
in IJJs, we want to provide explicit expressions for $c_N$ and
$c_1$ in terms of material parameters. Following the formalism of
Refs.~[\onlinecite{Fluxon}] and [\onlinecite{Modes}] we obtain
% *** EQUATION ***
\begin{eqnarray}
c_N &\simeq& \frac{c}{2}\sqrt{\frac{ts}{\varepsilon_\mathrm{r}
\lambda_{ab}^2}}\left[1+\frac18\left(\frac{s^2}{\lambda_{ab}^2}+\frac{\pi^2}{(N+1)^2}\right)\right],
\label{cnFull}\\
c_1 &\simeq& c\sqrt{\frac{ts}{\varepsilon_\mathrm{r}
\lambda_{ab}^2}}\left[\frac{s^2}{\lambda_{ab}^2}+\frac{\pi^2}{(N+1)^2}\right]^{-1/2}.
\label{c1Full}
\end{eqnarray}
% *** END EQUATION ***
The accuracy of expansion is
$\mathcal{O}[(s/\lambda_{ab})^4+(\pi/(N+1))^4]$. Here $c=3\times 10^8\,\mathrm{m/s}$
is the velocity of light in vacuum,
%$s=1.55\,\mathrm{nm}$ is the interlayer spacing in Bi-2212,
$\lambda_{ab}=\lambda_\mathrm{S}[s/d]^{1/2}$ is the effective
London penetration depth, and $t=s-d$ and $\varepsilon_\mathrm{r}$
are the thickness and the relative dielectric permittivity of the
junction barrier, respectively. The ratio
$t/\varepsilon_\mathrm{r} \simeq 0.1\,\mathrm{nm}$, corresponding
to reasonable values of the double-CuO layer thickness $d\simeq
0.6\,\mathrm{nm}$ and $\varepsilon_\mathrm{r} \simeq 10$, can be
deduced from previous studies of Fiske steps in
IJJs.\cite{Fiske,FiskeSlow,FFLatyshev} The first expansion term
$(s/\lambda_{ab})^2 \simeq 6\times 10^{-5}$ is small and can be
neglected in Eq.~(\ref{cnFull}) for all $N$ and in
Eq.~(\ref{c1Full}) for $N\ll\pi\lambda_{ab}/s \simeq 400$. This
leads to further simplification:
% *** EQUATION ***
\begin{eqnarray}
\label{cN}
c_N &\simeq& \frac{c}{2}\sqrt{\frac{ts}{\varepsilon_\mathrm{r} \lambda_{ab}^2}} = \frac{c_0}{\sqrt{2}} , \\
\label{c1}
c_1 &\simeq& c\sqrt{\frac{ts}{\varepsilon_\mathrm{r}\lambda_{ab}^2}}\frac{N+1}{\pi} \simeq \frac{2(N+1)}{\pi}c_N.
\end{eqnarray}
% *** END EQUATION ***
The accuracy of this expansion is $\mathcal{O}[(s/\lambda_{ab})^2+(N+1)^{-2}]$.

The above expressions allow straightforward estimation of extremal
velocities in IJJs. The slowest velocity $c_N \simeq (3.0 \pm 0.6)
\times 10^5\,\mathrm{m/s}$ is almost independent of $N$. Here, the
central value corresponds to optimal doping with $\lambda_{ab} \simeq 200~\mathrm{nm}$, and plus/minus
corrections to overdoped/underdoped Bi-2212 with smaller/larger
$\lambda_{ab}$, correspondingly. The fastest velocity $c_1$, by
contrast, is not universal and increases almost linearly with $N$.
For typical IJJ numbers used in previous studies, $c_1(N=30)\simeq
6\times 10^6\,\mathrm{m/s}$ and $c_1(N=100)\simeq 1.9\times
10^7\,\mathrm{m/s}$. Experimentally reported maximum flux-flow
velocities are of the order of
$c_N$\cite{Cherenkov,Fiske,FiskeSlow,Compar,Bae2007,Bae2009,FFLatyshev,FFlowBae,FFlowHatano}
and substantially smaller than $c_1$ for the corresponding number
of IJJs. According to numerical
simulations\cite{FFlowSimul,FFlowMachida} the rectangular lattice
is unconditionally stable only at fast superluminal velocities
$c_2<u_\mathrm{FF}<c_1$, which have not been reached in
experiments so far. Although the rectangular lattice can be
stabilized in the static case by geometrical confinement in small
mesas,\cite{Katterwe} it tends to reconstruct into the triangular
lattice at slow flux-flow velocities
$u_\mathrm{FF}<c_N$.\cite{FFlowSimul,FFlowMachida,FFlowKoshelev,FFlowRyndyk}
Therefore, the stability of the rectangular fluxon lattice is
remaining a critical issue for the realization of a coherent FFO.

At high in-plane magnetic fields $B>\Phi_0/s\lambda_\mathrm{J}$
fluxons form a regular lattice in the stack. This brings about
phonon-like collective excitations, which can be characterized by
two wave numbers\cite{KleinerModes} $k_m=m\pi/L ~(m=1,2,3,\ldots)$
in-plane and $q_n=n\pi/Ns ~(n=1,2,\ldots,N)$ in the $c$-axis
direction. Here $L$ is the in-plane length of the stack. Unlike
Josephson plasma waves, fluxon phonons have a linear dispersion
relation at low frequencies.\cite{Stephen} Fluxon phonons with
mode $(m,n)$ propagate with the in-plane velocity $c_n$ of
Eq.~(\ref{Cmodes}). The slowest velocity $c_N$ ($q_N=\pi/s$)
corresponds to out-of-phase oscillation in neighbor junctions, the
fastest $c_1$ ($q_1=\pi/Ns$) to in-phase oscillation of all
junctions. Fluxon phonons can be excited in the flux-flow state.
Geometric resonances occur when the ac-Josephson frequency
$\omega_\mathrm{J}=2 \pi V_\mathrm{FF}/\Phi_0$ coincides with the
frequency $\omega_{m,n}$ of one of the modes $(m,n)$. Here
% *** EQUATION ***
\begin{equation}\label{VFF}
V_\mathrm{FF}=u_\mathrm{FF}Bs
\end{equation}
% *** END EQUATION ***
is the dc flux-flow voltage per junction. The resonances can be
seen as a series of Fiske steps at voltages (per junction)
% *** EQUATION ***
\begin{equation}\label{Fiske}
V_{m,n}=\frac{\Phi_0}{2L} m c_n, \qquad({m=1,2,3,\ldots, ~n=1,2,\ldots,N}).
\end{equation}
% *** END EQUATION ***
The (almost) linear dispersion relation of fluxon phonons makes
Fiske steps with a given $n$ (almost) equidistant in voltage both
for single \cite{Kulik,Dmitr} and stacked\cite{SakUst} junctions.

From comparison of Eqs.~(\ref{VFF}) and (\ref{Fiske}) it follows
that modes with velocity $c_n$ can be excited at $u_\mathrm{FF} <
c_n$ provided that
% *** EQUATION ***
\begin{equation}\label{Mres}
\frac{u_\mathrm{FF}}{c_n} \geq \frac{\Phi_0}{2\Phi}.
\end{equation}
% *** END EQUATION ***
where $\Phi=BLs$ is the flux per junction. The strongest coupling
between resonant modes and flux-flow occurs when fluxons propagate
with the same velocity as fluxon phonons, $u_\mathrm{FF}=c_n$ (the
velocity matching condition) and the in-plane wavelength of the
standing wave is equal to the separation between fluxons.
This happens when
% *** EQUATION ***
\begin{equation}\label{VelMatch}
m^*= 2\Phi/\Phi_0.
\end{equation}
Therefore, the most prominent Fiske steps should correspond to the
velocity-matching modes $(m^*,n)$, even in the absence of Lorentz
contraction of the fluxon. Steps with odd and even $m$ should
modulate in anti-phase with each other,\cite{Kulik} and have
maximum amplitudes at integer and half-integer $\Phi/\Phi_0$,
respectively.\cite{Fiske,FiskeSlow} This additional selection rule
can make $m=m^*-1$ resonances stronger than $m=m^*$.

%******* TABLE *******
\begin{table*}[t]
\begin{ruledtabular}
 \begin{tabular}{cccdddddddcc}
%                                             Sven m4a     S23m2   S23m3   P22m4a          P22m4b
 $\mathrm{Mesa}$&
  Crystal&
  $N$&
  \multicolumn{1}{c}{$T_\mathrm{c}~(\mathrm{K})$}&
  \multicolumn{1}{c}{$L~({\muup \mathrm{m}})$}&
  \multicolumn{1}{c}{$w~({\muup \mathrm{m}})$}&
  \multicolumn{1}{c}{$L_\mathrm{eff}~({\muup\mathrm{m}})$}&
  \multicolumn{1}{c}{$B_0~({\mathrm{T}})$}&
  \multicolumn{1}{c}{$J_\mathrm{c}~(\mathrm{kA/cm^2})$}&
  \multicolumn{1}{c}{$\lambda_\mathrm{J}~(\muup {\mathrm{m}})$}&
  $V_{1,N}~({\mathrm{mV}})$&
  $c_N~({10^5\,\mathrm{m/s}})$\\
  \hline
  $1$& Bi-2212    & $8 $&         82 & 2.7 & 1.4 & 2.43 & 0.55 & 1.1  & 0.69 & $0.135\pm 0.005$&    $3.17\pm 0.15$\\ %    Sven m4a
  $2$& Bi-2212    & $11$&         89 & 1.0 & 1.0 & 1.60 & 0.82 & 1.1  & 0.67 & $0.185\pm 0.010$&    $2.88\pm 0.20$\\ % S23m2
  $3$& Bi(Pb)-2212 & $\approx 15$      & 87 & 0.6 & 1.1 & 0.53 & 2.54 & 11.5 & 0.20 & $0.64 \pm 0.04$ &    $3.3 \pm 0.2$\\ % P22m4b
  $4$& Bi(Pb)-2212 & $\approx 56$& 87 & 1.2 & 2.0 & 1.13 & 1.21 & 7.3  & 0.26 & $0.38 \pm 0.02$ &    $3.9 \pm 0.2$\\ % P22m4a
 \end{tabular}
\end{ruledtabular}
\caption{\label{tab:sizedependence}
Summary of  studied mesas. $N$ is the number of junctions in the mesa. $L$ and $w$ are the
nominal in-plane mesa sizes in directions perpendicular and parallel to the field, respectively. $L_\mathrm{eff}=\Phi_0/B_0s$ is the effective length calculated from the observed flux-quantization field $B_0$. $J_\mathrm{c}$ is the
critical current density at $B=0$. $V_{1,N}$ is the voltage of the lowest out-of-phase Fiske
step and $c_N$ is the corresponding slowest velocity of
electromagnetic waves. The pure and lead-doped Bi-2212 crystals are slightly underdoped and overdoped, respectively.}
\end{table*}
%******* END TABLE *******
\section{Experimental}
%**********************
\label{sec:experimental} Two batches of Bi-2212 single crystals
were used in this work: pure Bi-2212 and lead-doped
$\mathrm{Bi_{2-y}Pb_{y}Sr_2CaCu_2O}_{8+x}$ [Bi(Pb)-2212] single
crystals. The most noticeable difference between those crystals is
in the $c-$axis critical current density $J_\mathrm{c}$, which is
larger by an order of magnitude for Bi(Pb)-2212, see
Table~\ref{tab:sizedependence}. Comparing with the doping
dependence of $J_\mathrm{c}$ in IJJs\cite{Doping} we conclude that
Bi-2212 crystals are slightly underdoped and Bi(Pb)-2212 slightly
overdoped.

Small mesa structures were fabricated at the surface of freshly
cleaved crystals by means of photolithography, Ar milling and
focused ion beam trimming. Details of mesa fabrication are
described elsewhere.\cite{KrSubmicron} Several mesas with
different dimensions were studied, all of them showed similar
behavior. The number of junctions in the mesas $N$ was obtained by
counting quasiparticle (QP) branches at $B=0$, see Fig.
\ref{fig:Fig1}. Mesa properties are summarized in
Table~\ref{tab:sizedependence}.

Samples were mounted on a rotatable sample holder and carefully
aligned to have $B\, \|\ ab$. %perpendicular to one side of the mesa.
Accurate alignment is critical for the resonant phenomena reported
below. To avoid field lock-in hysteresis, the alignment was done
by minimizing the high field and high bias QP
resistance.\cite{Katterwe} After alignment all mesas exhibited
clear Fraunhofer-like modulation of the critical
current\cite{Katterwe} $I_\mathrm{c}(B)$ with the flux
quantization field $B_0$, see Table~\ref{tab:sizedependence}. This
allows estimation of the actual junction length, perpendicular to
the field, $L_\mathrm{eff}=\Phi_0/B_0s$. It is usually only
slightly different from the nominal mesa size $L$ obtained from
the surface inspection by SEM. Fiske steps were observed in all
studied mesas after proper alignment. Measured values of the
lowest Fiske step $V_{1,N}$ and the corresponding lowest
velocities $c_N$ are given in Table~\ref{tab:sizedependence}. Step
voltages are inversely proportional to mesa lengths, as expected
from Eq.~(\ref{Fiske}).

All measurements were performed at $T=1.6\,\mathrm{K}$. $I$-$V$
characteristics are presented as digital oscillograms (intensity
plots) and were obtained by measuring voltage and current over
several current sweeps. The contact resistance originating from
the deteriorated topmost IJJ\cite{SecondOrder} was numerically
subtracted from all $I$-$V$ characteristics.\cite{Katterwe}
Subtraction works very well in the whole flux-flow region with the
accuracy $\sim 10\,\mathrm{\muup V}$ (cf. supercurrent branches
in the figures below). It simplifies data analysis and does not
affect results in any way.

\section{Results and Discussion}
%**********************
\label{sec:resultsDiscussion}
%**********************
\subsection{Flux flow and Fiske steps}
\begin{figure}[t]
   \includegraphics[width=0.45\textwidth]{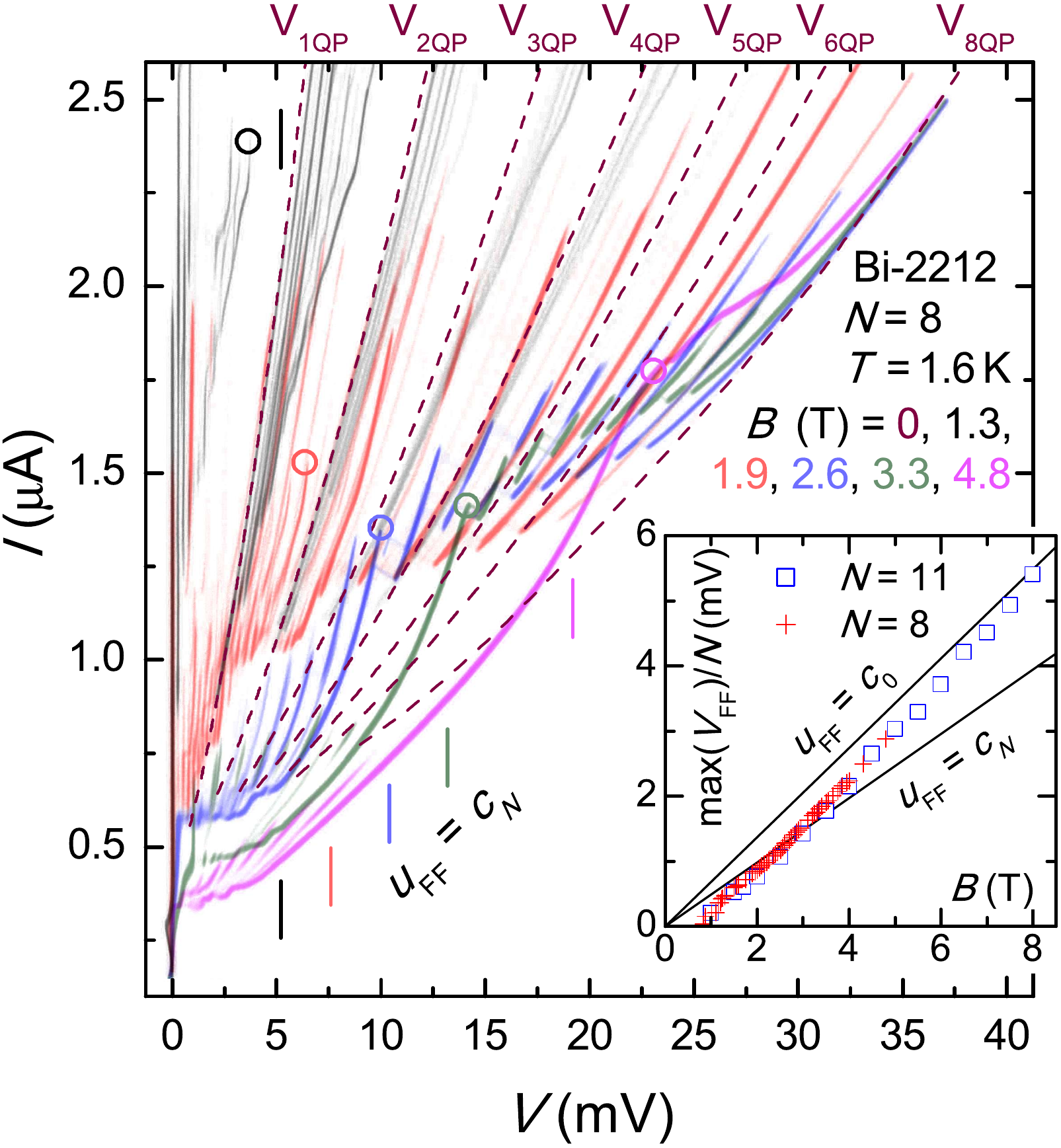}
    \caption{\label{fig:Fig1}(Color online)
$I$-$V$ characteristics of mesa~1 %with $N=8$ IJJs
at different magnetic fields. Dashed lines represent quasiparticle
branches at $B=0$. Development of flux-flow branches with
increasing field is clearly seen. Circles mark maximum flux-flow
voltages. Short lines indicate the velocity matching condition
$u_\mathrm{FF}=c_8$. Inset: The maximum flux-flow voltage per
junction vs. $B$ for mesa~1 ($N=8$) and mesa~2 ($N=11$). Lines
correspond to flux-flow velocities equal to $c_N$ and $c_0$,
respectively. It is seen that the experimentally observed
flux-flow velocity becomes superluminal, $u_\mathrm{FF}>c_N$, at
$B>2.6$T and approaches the Swihart velocity $c_0$ at high
fields.}
\end{figure}

%**********************
Figure~\ref{fig:Fig1} shows $I$-$V$ oscillograms of mesa~1 at
different magnetic fields from $0$ to $4.8\,\mathrm{T}$. At
$I>I_\mathrm{c}$, fluxons start to move and soon all 8 IJJs are in
the flux-flow state with a total voltage $V_\mathrm{8FF}\sim
8V_\mathrm{FF}$, where $V_\mathrm{FF}$ is the flux-flow voltage
per junction. When one of the junctions switches to the QP state,
$7$ junctions remain in the flux-flow state and the first combined
QP-FF branch appears at $V=V_\mathrm{1QP}+V_\mathrm{7FF}$. When
the next IJJ switches to the QP state the second combined QP-FF
branch appears at $V=V_\mathrm{2QP}+V_\mathrm{6FF}$, then at
$V=V_\mathrm{3QP}+V_\mathrm{5FF}$, and so on.\cite{Irie} For
identical junctions $V_{i\mathrm{QP}}= iV_\mathrm{QP}$, and
$V_{i\mathrm{FF}}\approx iV_\mathrm{FF}$. %when all junctions participate in the flux flow.
The separation between such combined QP-FF branches is
$V_\mathrm{QP}-V_\mathrm{FF}$ and is decreasing with increasing
field to finally disappear at $B \sim 5.8\,\mathrm{T}$ (not
shown). This is consistent with previous
observations.\cite{Bae2007} What is different is the presence of
remarkably strong Fiske steps at every QP-FF branch.

\begin{figure*}[t]
\begin{center}
 \begin{tabular}{@{} cp{4mm}c @{}}
   \includegraphics[height=0.45\textwidth]{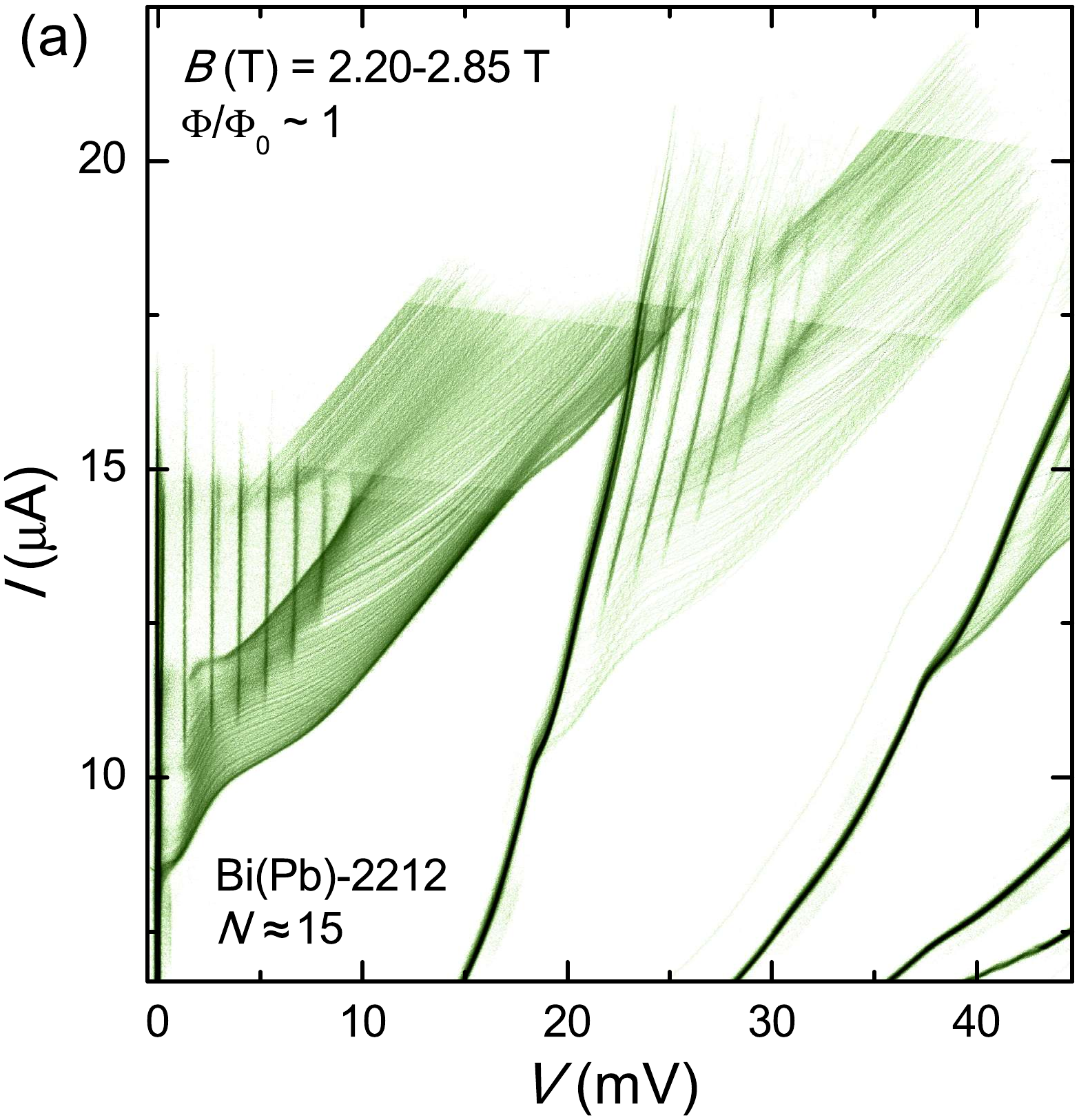} && \includegraphics[height=0.45\textwidth]{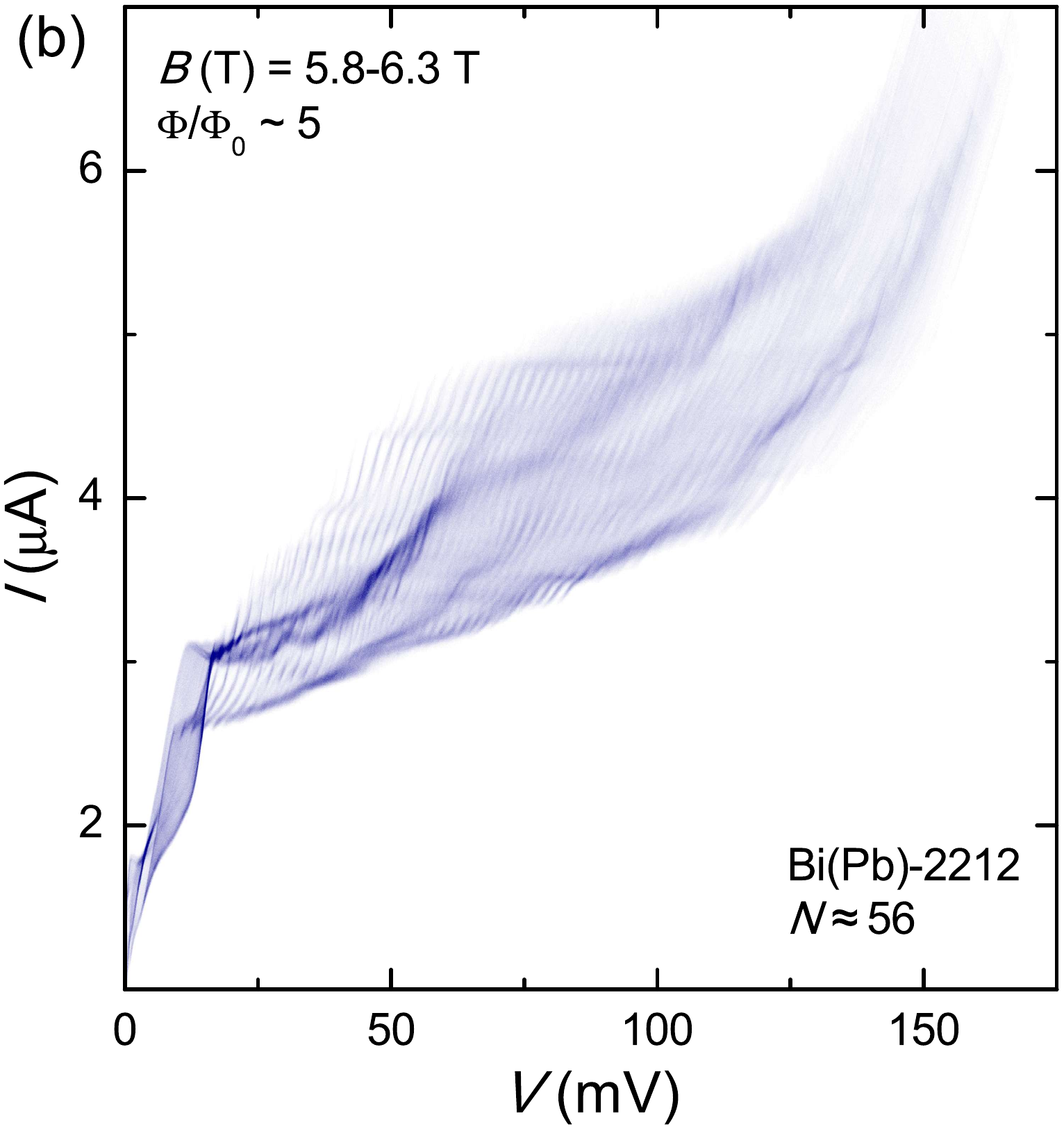} \\
 \end{tabular}
\end{center}
\caption{\label{fig:Fig2} (Color online) $I$-$V$ oscillograms for
continuous field sweeps around integer flux quanta per junction
for Bi(Pb)-2212 mesas (a)~3 and (b)~4. At low fields in (a) both
the continuously varying viscous flux-flow state and sharp,
vertical, equidistant Fiske steps are simultaneously observed. At
higher fields in (b) the whole flux-flow oscillogram is stratified
into dense grids of resonances. This clearly demonstrates that the
flux-flow $I$-$V$ characteristics in this case are not changing
continuously with field, but switch between a variety of distinct
resonant states.}
\end{figure*}

Figure~\ref{fig:Fig2} shows oscillograms collected during
continuous field sweeps for Bi(Pb)-2212 mesas 3 and 4. It is
clearly seen that the flux-flow characteristics are not continuous
but are composed of a sequence of distinct Fiske steps, much like
the case of strongly underdamped low-$T_\mathrm{c}$
junctions.\cite{Koshelets} We emphasize that such high-quality
geometrical resonances are observed only after careful alignment
of the sample. Even a small misalignment of $\sim 0.1^\circ$ leads
to an avalanche-like entrance of Abrikosov vortices at high
fields, which completely suppresses both these spectacular Fiske
steps and the Fraunhofer modulation of the critical
current.\cite{Katterwe}

%******* FIGURE *******
%******* END FIGURE *******

Unambiguous identification of the resonance modes $(m,n)$ requires
first of all discrimination between individual (not phase-locked)
and collective (phase-locked) Fiske steps. This is nontrivial,
because, for instance, the voltage $1V_{m,1}$ of an individual
in-phase step is close to the voltage $NV_{m,N}$ of a collective
out-of-phase step, see Eqs.~(\ref{cN},\ref{c1}). Besides, not all
junctions may participate in the collective resonance. For mesas
with large $N>10$ different resonant modes $(m,n)$ form a very
dense, almost continuous, Fiske step sequence. Therefore, bare
analysis of voltage spacings is insufficient. To simplify
discrimination between modes with different velocities, mesas with
small $N$ should be analyzed. Below we mainly focus on mesa~1 with
the smallest number of IJJs $N=8$.

%******* FIGURE *******
%******* END FIGURE *******

\begin{figure*}[t]
\begin{center}
 \begin{tabular}{@{} cp{4mm}c @{}}
   \includegraphics[height=0.45\textwidth]{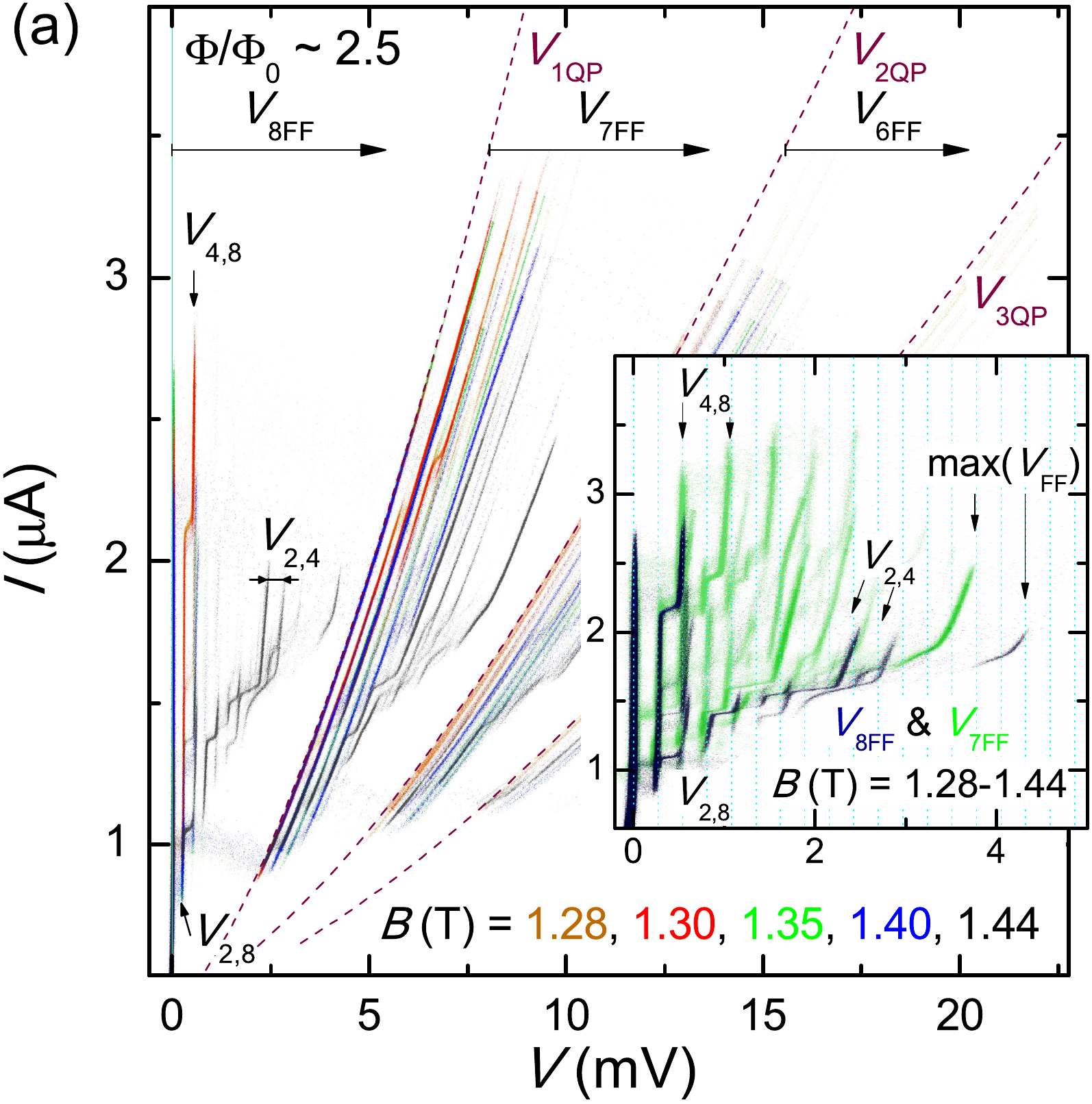} && \includegraphics[height=0.45\textwidth]{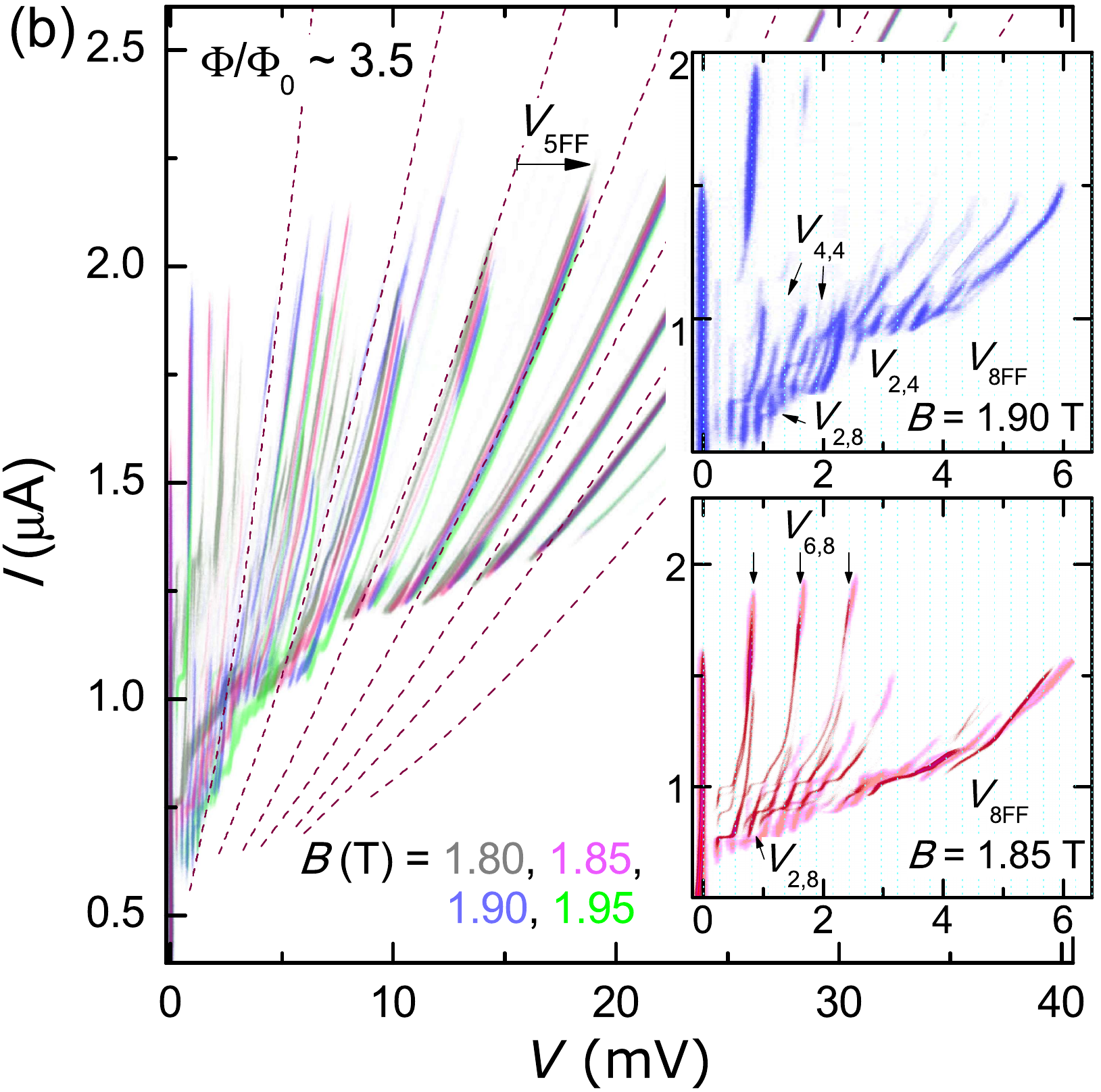} \\
\\
   \includegraphics[height=0.45\textwidth]{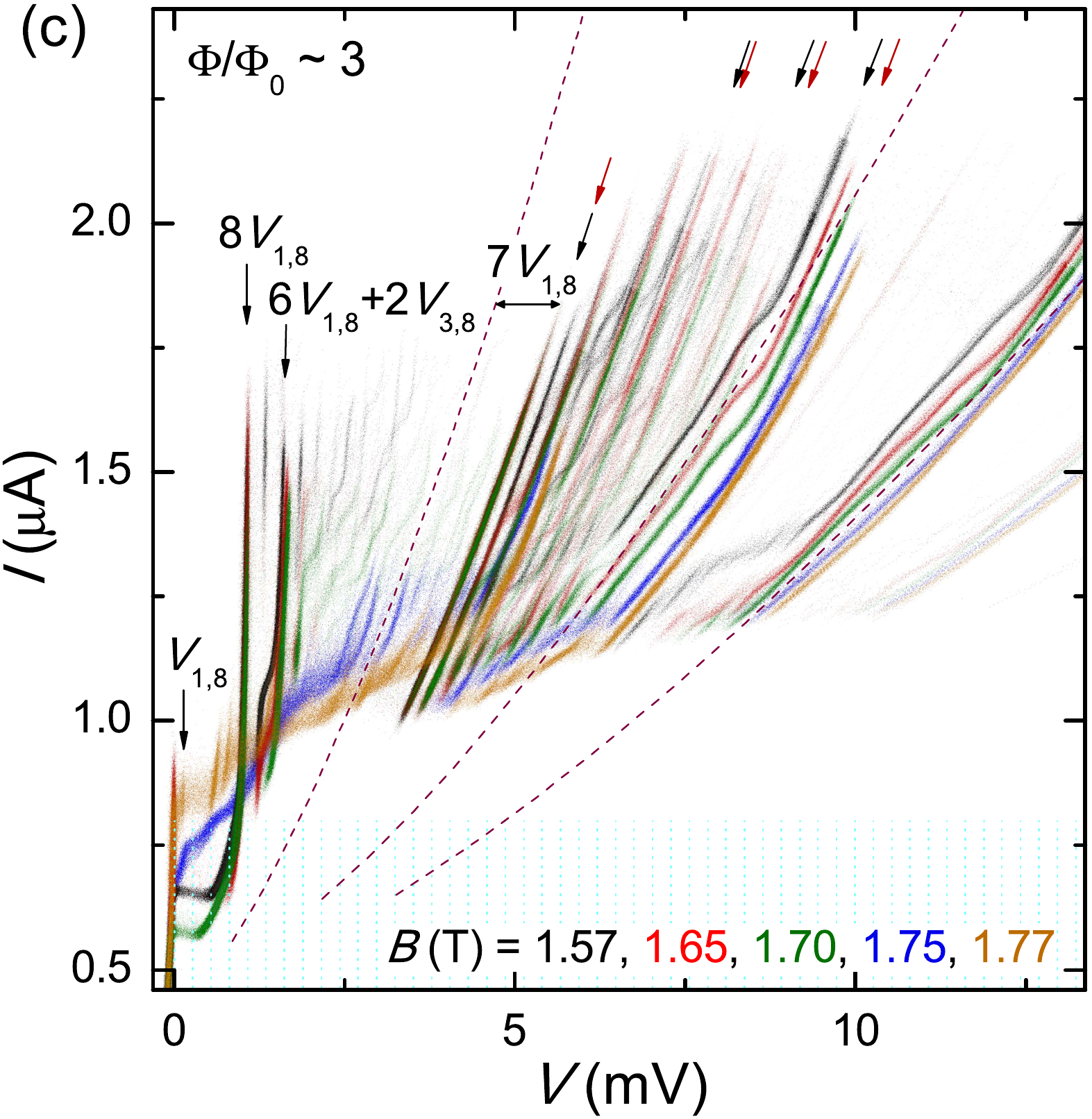} && \includegraphics[height=0.45\textwidth]{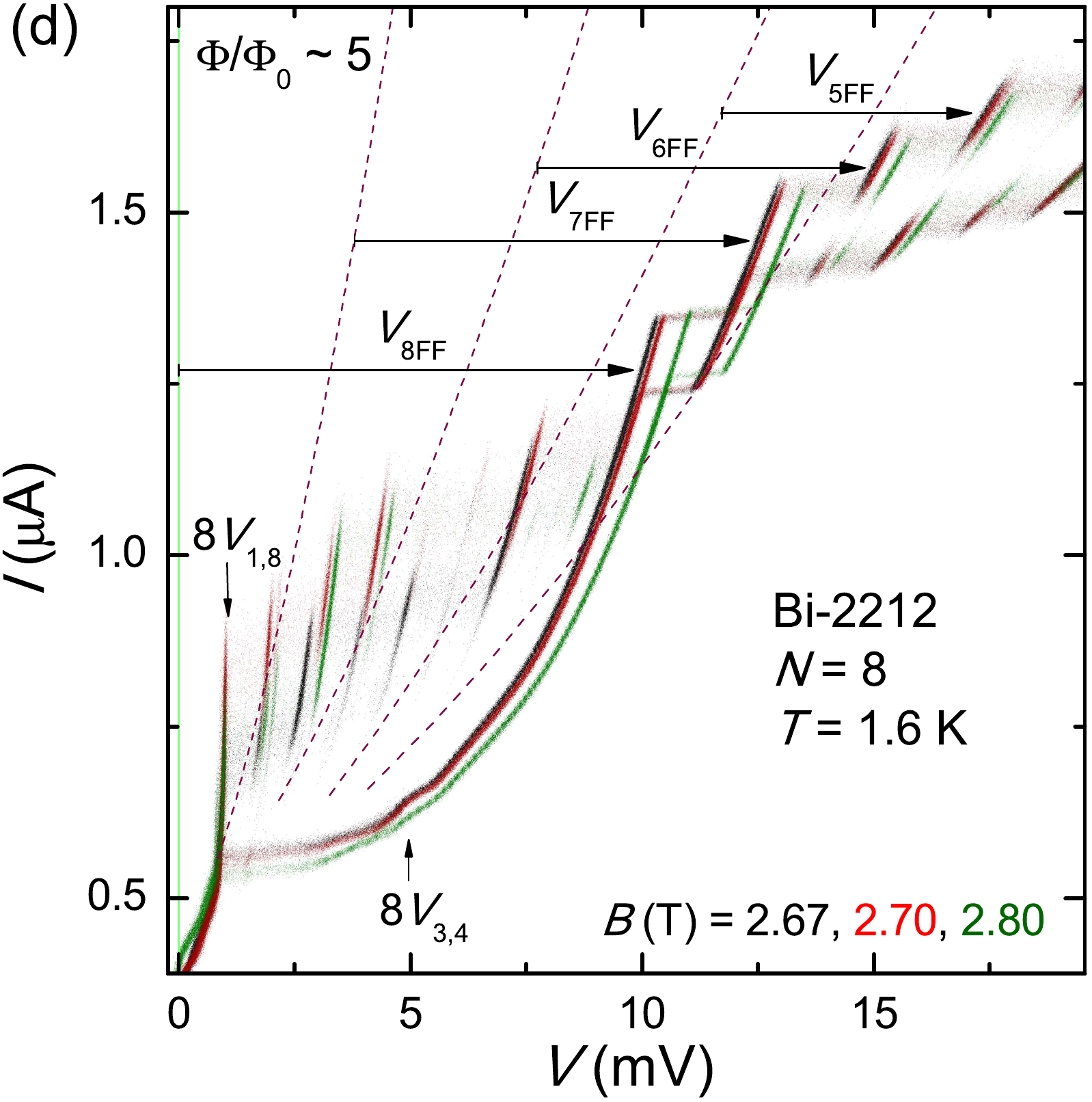} \\
 \end{tabular}
\end{center}
\caption{\label{fig:Fig3} (Color online) $I$-$V$ oscillograms of
mesa~1 (at constant $B$) in narrow field ranges near half-integer
[(a) and (b)] and integer [(c) and (d)] $\Phi/\Phi_0$. A large
variety of high-quality Fiske steps are seen. Also shown are the
QP branches at $B=0$ (dashed lines). The grid-spacing is
$V_{2,8}=0.27\,\mathrm{mV}$. The inset in (a) shows Fiske steps at
the flux-flow branches with zero ($V_{8\mathrm{FF}}$) and one
($V_{7\mathrm{FF}}$) IJJ in the QP state, respectively. Voltage
separation between steps is clearly independent of the number of
junctions, showing that the Fiske steps here are from individual
IJJs. The insets of panel~(b) show the low bias region at
$B=1.85\,\mathrm{T}$ and $1.90\,\mathrm{T}$. Several individual
Fiske steps are marked. At integer $\Phi/\Phi_0$ in (c), by
contrast, the prominent collective step $NV_{1,8}$ is seen, which
scales with the number of junctions for $V_{8\mathrm{FF}}$ and
$V_{7\mathrm{FF}}$. Two collective steps are also indicated in
(d). Thus, IJJs behave more individually or collectively at
half-integer or integer $\Phi/\Phi_0$, respectively.}
\end{figure*}

Figure~\ref{fig:Fig3} shows $I$-$V$ oscillograms of mesa~1 in
narrow field ranges corresponding to half-integer and integer
number of flux quanta $\Phi/\Phi_0$ per IJJ. To check whether the
observed Fiske steps are individual or collective, in the inset of
Fig.~\ref{fig:Fig3}(a) we compare steps at the first
$V_{8\mathrm{FF}}$ flux-flow branch and the second
$V_{7\mathrm{FF}}$ (with one junction in the QP state). To obtain
$V_{7\mathrm{FF}}$, the first QP voltage $V_{1\mathrm{QP}}(B=0)$
is subtracted from the $V_{1\mathrm{QP}}+V_{7\mathrm{FF}}$ branch.
Collective steps should decrease in proportion to the number of
junctions, e.g., $7/8$ times. Individual steps, on the other hand,
should maintain the same voltage at both flux-flow branches. This
is the case at half-integer $\Phi/\Phi_0$ in
Fig.~\ref{fig:Fig3}(a) and (b). The dominant step at integer
$\Phi/\Phi_0\simeq 3$, see Fig.~\ref{fig:Fig3}(c), however, occurs
at $8V_{1,8}$ at the $V_{8\mathrm{FF}}$ branch and at $7V_{1,8}$
at the next $V_{7\mathrm{FF}}$ branch, and is thus collective.
Discrimination between individual and collective steps allows
accurate identification of resonant modes $(m,n)$.

%******* FIGURE *******
%******* END FIGURE *******

The smallest step sequence $\simeq 0.135\,\mathrm{mV}$ is observed
at integer $\Phi/\Phi_0$, see Fig.~\ref{fig:Fig3}(c). It
corresponds to the expected value $V_{1,8}$ for the lowest odd
$m=1$ out-of-phase $n=8$ individual Fiske step. The corresponding
slowest velocity $c_8\simeq 3.17 \times 10^5\,\mathrm{m/s}$ is
consistent with the estimation from
Sec.~\ref{sec:generalRelations}. Simultaneously, a strong
collective Fiske step at $V=8V_{1,8}$ is observed, as marked in
Figs.~\ref{fig:Fig3}(c) and (d). It corresponds to the
phase-locked state of the whole mesa at the same $(1,8)$
resonance. Steps with separation $2V_{1,8}$ are seen above the
collective step in Fig.~\ref{fig:Fig3}(c). They are due to
switching between nearest odd $m=1$ and 3 out-of-phase $n=8$
resonances in individual IJJs, consistent with numerical
simulations below.

At half-integer $\Phi/\Phi_0$ the smallest step sequence
corresponds to the lowest even $m=2$ individual Fiske step
$V_{2,8} \simeq 0.27\,\mathrm{mV}$, marked in
Fig.~\ref{fig:Fig3}(a) and insets of Fig.~\ref{fig:Fig3}(b). The
most prominent Fiske steps at half-integer $\Phi/\Phi_0$
correspond to individual even $m=4$ and $m=6$ out-of-phase
resonances  in Figs.~\ref{fig:Fig3}(a) and \ref{fig:Fig3}(b),
respectively, which are close to the velocity matching condition
of Eq.~(\ref{VelMatch}).

While the Fiske step patterns vary significantly with magnetic
field, they have certain common features for integer (more
collective behavior, odd-$m$ modes) and half-integer (more
individual behavior, even-$m$ modes) $\Phi/\Phi_0$. The step
amplitudes oscillate strongly with $B$, with odd and even $m$ in
anti-phase with each other, as expected. A complete overview of
the modulation can be found in the supplementary.\cite{Supplem}

%******* TABLE *******
\begin{table}[t]
   \begin{ruledtabular}
       \begin{tabular}{ccccccccc}
       $n$&1&2&3&4&5&6&7&8\\
       \hline
       $c_n\,({10^5\,\mathrm{m/s}})$ &17.97&9.13 &6.24 &4.86 &4.08 &3.60 &3.32 &3.17\\
       $c_n/c_8$                              &~5.67&2.88 &1.97 &1.53 &1.29 &1.14 &1.05 &1.00\\
       $V_{1,n}\,({\mathrm{mV}})$    &0.765&0.389&0.266&0.207&0.174&0.153&0.141&0.135\\
       \end{tabular}
       \end{ruledtabular}
   \caption{\label{tab:velocities}
   Calculated characteristic velocities $c_n$ for mesa~1, using $s=1.55\,\mathrm{nm}$, $\lambda_{ab}=200\,\mathrm{nm}$, and
   $t/\varepsilon_\mathrm{r} =0.112\,\mathrm{nm}$ and Eq. (\ref{Cmodes}). Also given are
the ratios $c_n/c_8$ and calculated Fiske step voltages $V_{1,n}=
B_0 s c_n/2$, where $B_0=0.55\,\mathrm{T}$ is the
flux-quantization field.}
\end{table}
%******* END TABLE *******

\subsection{Superluminal resonances}
%**********************
So far, we discussed out-of-phase Fiske steps with voltages equal
to multiple integers of $V_{1,8}$. Such steps, both
individual\cite{FiskeInd} and collective \cite{Fiske,FiskeSlow},
with similar velocities $c_N=2.5-3.5 \times 10^5\,\mathrm{m/s}$
were observed previously, although with smaller relative
amplitudes. However, in our case the full set of observed Fiske
steps cannot be described solely by multiple integers of
$V_{1,8}$. We can distinguish step sequences with different
voltage spacings, which must correspond to resonances with higher
velocities $c_n>c_8$, listed in Table~\ref{tab:velocities}.

In the flux flow region $V_{7\mathrm{FF}}$ of
Fig.~\ref{fig:Fig3}(c), where one IJJ has switched to the QP
state, two sequences of steps with slightly different voltage
spacings are seen at $1.57\,\mathrm{T}$ (black) and
$1.65\,\mathrm{T}$ (red curve). Both step sequences start from the
same $7V_{1,8}$ level, corresponding to the collective $(1,8)$
geometrical resonance in the remaining 7 IJJs. The lowest (black)
step sequence has a step separation of $4V_{1,8}$ and is due to
sequential switching of junctions from the lowest odd $m=1$
resonance $V_{1,8}$ to the odd resonance $V_{5,8}$ closest to the
velocity matching condition at the same speed. The whole step
sequence is then described by $V_i=(7-i)V_{1,8}+iV_{5,8}$, where
$i=0, 1,\,\ldots, 7$ is the number of junctions in the $V_{5,8}$
state and the separation between steps is
$V_{5,8}-V_{1,8}=4V_{1,8}$. The higher (red) step sequence also
starts from the same $7V_{1,8}$ level, but the splitting between
the two step sequences becomes progressively larger with
increasing step number, as indicated by black and red arrows for
steps number $3$ and $5$ in Fig.~\ref{fig:Fig3}(c). This step
sequence is described by a similar expression
$V_i=(7-i)V_{1,8}+iV_{m,n}$, ($i=0, 1,\,\ldots, 7$). The
corresponding eight (red) Fiske steps can be distinguished in
Fig.~\ref{fig:Fig3}(c). The involved resonance $V_{m,n}= 1.05
V_{5,8}$ is very close to $V_{5,7}$, which represents the first
superluminal resonance $n=7$, see Table~\ref{tab:velocities}.

Individual step sequences
\begin{equation}\label{Vi}\nonumber
V_i=(N-i)V_{1,8}+iV_{m,n},\qquad({i=0, 1,\,\ldots, N})
\end{equation}
are observed at all integer $\Phi/\Phi_0$. At $\Phi/\Phi_0\simeq
5$ in Fig.~\ref{fig:Fig3}(d), for example, nine steps can be
distinguished on the $8V_\mathrm{FF}$ branch. At a first glance
they may look similar to previously seen sub-branching due to
onset of uncorrelated viscous flux-flow in individual
IJJs.\cite{Bae2007} However, steps discussed here are resonant. In
contrast to viscous flux-flow branches, which change monotonously
with field, see Fig.\ref{fig:Fig2}(a), these steps are
periodically modulated in magnetic field\cite{Supplem} and the
step voltage is not changing continuously, but goes through
discrete, although dense, set of values, see Fig.
\ref{fig:Fig2}(b). This leads to the small splitting of steps in
Fig.~\ref{fig:Fig3}(c). Splitting of steps number $2$~--~$4$ is
also seen in Fig.~\ref{fig:Fig3}(d). At certain conditions
different step sequences can be observed at the same $B$, leading
to a fine splitting as shown in the bottom inset of
Fig.~\ref{fig:Fig3}(b). The appearance of different step sequences
is due to discrete, rather than continuous, change of the
flux-flow voltage with $B$.

At $\Phi/\Phi_0\sim 2.5$, see Fig.~\ref{fig:Fig3}(a), an
individual step sequence with a voltage spacing of
$0.42\,\mathrm{mV}$ is observed both in the flux-flow regions
$V_{8\mathrm{FF}}$ and $V_{7\mathrm{FF}}$. The ratio of this
voltage spacing to $V_{2,8}$ is $1.55$, which is close to the
ratio expected for $c_4/c_8=1.53$, see Table~\ref{tab:velocities}.
We, therefore, identify this voltage spacing with $V_{2,4}$. At
the next half-integer $\Phi/\Phi_0\sim 3.5$ we observe two
even-$m$ resonances for the same $n=4$ mode, $V_{2,4}$ and
$V_{4,4}$, as indicated in the top inset of
Fig.~\ref{fig:Fig3}(b). In Fig.~\ref{fig:Fig3}(d), an additional
collective step is seen at $V\simeq 5\,\mathrm{mV}$ when the
current is decreased from high bias. It may correspond to a
phase-locked $8V_{3,4}$ Fiske step, or possibly $8V_{4,6}$.

\subsection{Fiske step modulation}
\begin{figure}[t]
\includegraphics[width=0.5\textwidth]{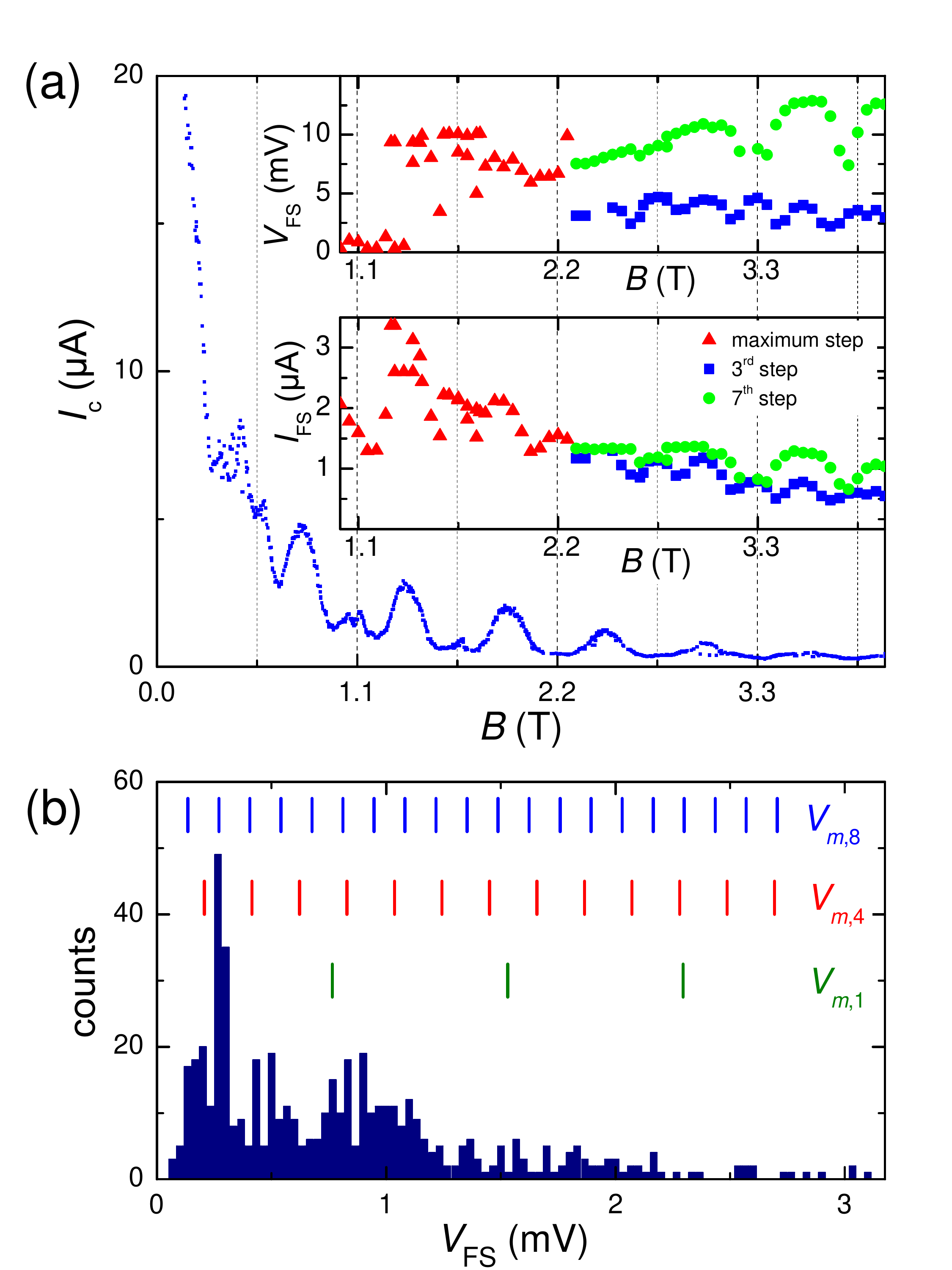}
\caption{\label{fig:Fig4}(Color online) (a) $I_\mathrm{c}(B)$
pattern for mesa~1. Grid lines corresponds to integer
$\Phi/\Phi_0$. The lower inset shows modulation of amplitudes of
some prominent Fiske steps. Triangles represent the Fiske steps
with the highest amplitude at fields from $1\,\mathrm{T}$ to
$2.25\,\mathrm{T}$. Squares and circles represent the
$3^{\mathrm{rd}}$ resp. $7^{\mathrm{th}}$ step in individual Fiske
step sequences at high bias at $B>2.3\,\mathrm{T}$ [as in
Fig.~\ref{fig:Fig3}(d) at $0.5\,\mathrm{\muup A} < I < 1.5
\,\mathrm{\muup A}$]. Voltages of the same steps are shown in the
upper inset. (b) Probability histogram for observation of
different steps separations in the field interval from 0 to
$4\,\mathrm{T}$. Top vertical lines indicate out-of-phase Fiske
steps, which are multiple integers of $V_{1,8}$. Apparently all
observed steps cannot be described solely by out-of-phase
resonances. Expected voltages for $n=4$ and $n=1$ superluminal
steps are also indicated.}
\end{figure}

%**********************
The main panel in Fig.~\ref{fig:Fig4}(a) shows Fraunhofer-like
modulation of the critical current for mesa~1. Dashed vertical
lines indicate the magnetic fields at integer $\Phi/\Phi_0$.
Dominant and sub-dominant maxima of $I_\mathrm{c}(\Phi)$ at
half-integer and integer $\Phi/\Phi_0$ correspond to rectangular
and triangular fluxon lattices, respectively.\cite{Katterwe} Fiske
steps also modulate strongly with magnetic field, as shown in the
lower inset of Fig.~\ref{fig:Fig4}(a) for several prominent steps.
It is expected that steps with even $m$ modulate in-phase with
$I_\mathrm{c}(\Phi)$, i.e., have maxima at half-integer
$\Phi/\Phi_0$. Steps with odd $m$ modulate out-of-phase with
$I_\mathrm{c} (\Phi)$ and have maxima at integer $\Phi/\Phi_0$.
Voltages of the chosen prominent Fiske steps are shown in the
upper inset. Variation of the voltage demonstrates that switching
between modes with different $m$ and $n$ occurs with variation of
field.

%******* FIGURE *******
%******* END FIGURE *******

Figure~\ref{fig:Fig4}(b) summarizes our main experimental results.
It shows a probability histogram for all observed step voltage
spacings in the field interval up to $B=4\,\mathrm{T}$. The step
at $V_{2,8}\simeq 0.27\,\mathrm{mV}$ has the highest number of
counts. However, vertical lines in the upper part of the figure
indicate that all experimentally observed steps cannot be
described solely by $(m,8)$ resonances. To explain the rich
variety of steps, geometrical resonances with different velocities
have to be present. Exact identification of all steps is rather
complicated because velocity splitting creates rather dense and
often overlapping voltage sequences even for $N=8$. Expected
voltage positions for individual superluminal $n=4$ and $n=1$
Fiske steps are indicated in Fig.~\ref{fig:Fig4}(b).

\subsection{The maximum flux-flow velocity}
%**********************
\label{sec:FluxFlowVelocity} Since we have confidently identified
the slowest velocity of light $c_8=3.17 \times
10^5\,\mathrm{m/s}$, it is instructive to compare it with the
maximum flux-flow velocity. In Fig.~\ref{fig:Fig1} voltages at the
velocity matching condition $u_\mathrm{FF}=c_8$ and maximum
flux-flow voltages are marked by short lines and circles,
respectively. It is seen that the maximum flux-flow velocity is
less, equal, and larger than $c_8$ for fields less, equal, and
larger than $2.6\,\mathrm{T}$. Interestingly, nothing special
happens with the flux-flow branch when it becomes superluminal,
$u_\mathrm{FF}>c_N$. This is consistent with numerical
simulations\cite{Fluxon} and is due to the absence of Lorentz
contraction of fluxons in strongly coupled, stacked Josephson
junctions. The inset in Fig.~\ref{fig:Fig1} shows measured maximum
flux-flow voltages per junction for mesas~1 and 2. The lines
represent flux-flow voltages with velocities $c_N=3.17\times
10^5\,\mathrm{m/s}$ and $c_0=4.42\times 10^5\,\mathrm{m/s}$.
Gradual increase of the maximum flux-flow velocity with field is
clearly seen and is consistent with previous
reports.\cite{Fiske,Compar,Cherenkov,FFlowHatano,FFLatyshev} The
maximum $u_\mathrm{FF}$ is approaching $c_0$ at high fields. A
similar limiting velocity was obtained by numerical simulations
for the case when fluxon stability is limited by Cherenkov
radiation (see Fig. 7 in Ref. \onlinecite{Fluxon}).

\section{Numerical modelling}
%**********************
\label{sec:numericalModelling} To get a better insight into
experimental data, we performed numerical simulations of the
coupled sine-Gordon equation with non-radiating boundary
conditions (see Refs.~[\onlinecite{Fluxon,Modes}] for details of
the numerical procedure, analysis of Fiske steps with proper
radiative boundary conditions can be found in
Ref.~[\onlinecite{TheoryFiske}]). To reproduce the real situation,
we introduced a small gradient of $I_\mathrm{c}$ ($5\%$ per IJJ)
from the top to the bottom of the mesa. We also considered the
presence of the bulk crystal below the mesa and the additional top
deteriorated junction.\cite{SecondOrder} To model the crystal, we
introduced an additional junction at the bottom of the stack with
zero net bias current. The additional top junction was assumed to
have a 100 times smaller $I_\mathrm{c}$. Passive top and bottom
junctions do not cause any principle differences in fluxon
dynamics. Although they may participate in geometrical resonances,
they do not contribute to the flux-flow voltage. In simulations we
used parameters of the mesa 1: $N=8$, $L=3.5\lambda_\mathrm{J}
\simeq 2.4\,\mathrm{\muup m}$, and the damping coefficient $\alpha
=0.01-0.05$. Small damping is important for excitation of a large
variety of geometrical resonances. In order to reach various
resonances the bias current was swept back and forth several
times, mimicking experimental measurement.

\begin{figure}[t]
\includegraphics[width=0.49\textwidth]{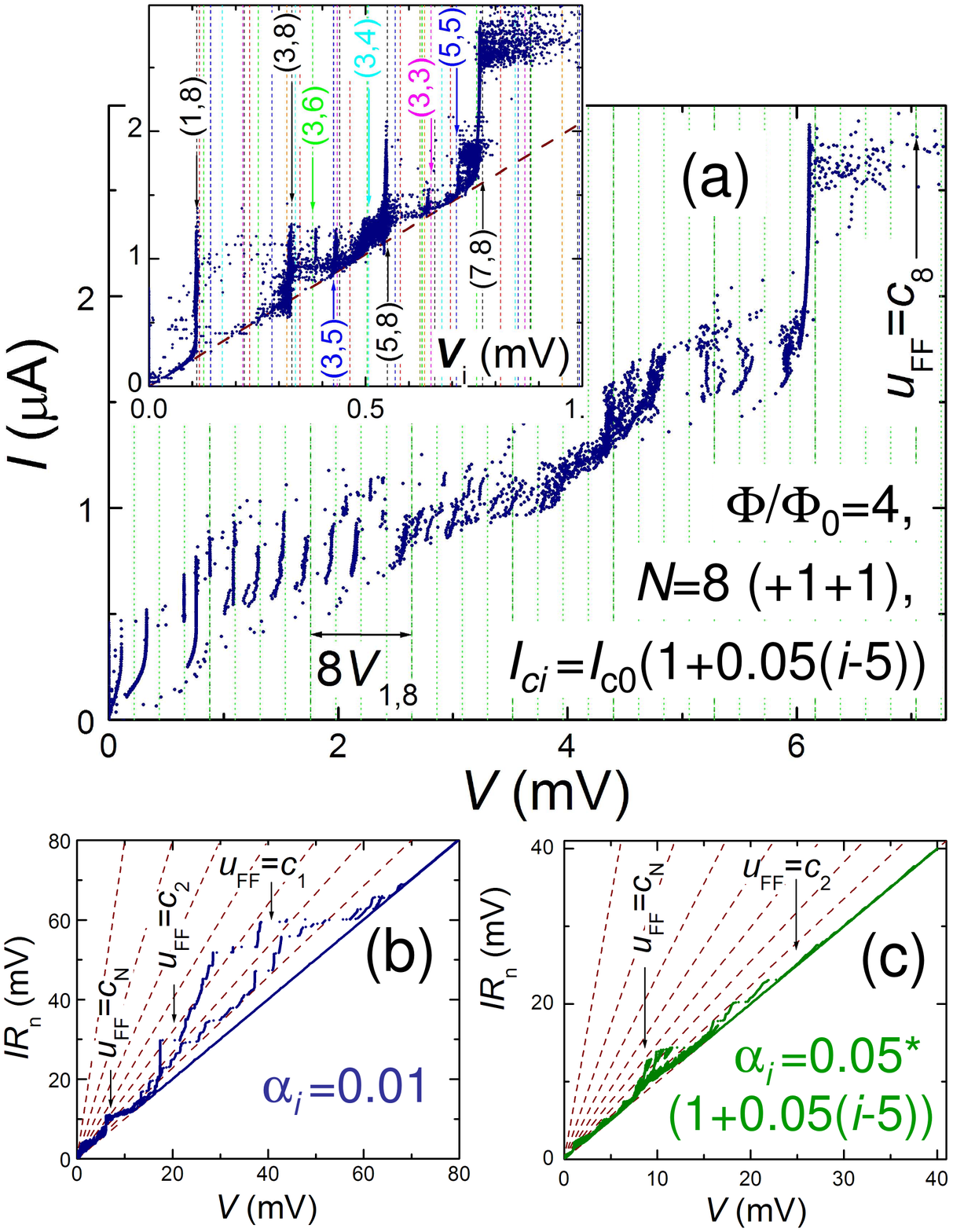}
\caption{\label{fig:Fig5}(Color online) Simulated $I$-$V$
characteristics at $\Phi/\Phi_0=4.0$ for a mesa with $N=8$ active
IJJs and two passive junctions at the top and bottom of the mesa.
Panels (a) and (b) represent simulations for a small nonuniformity
of the critical current ($5\%$ per IJJ) and uniform damping
parameter $\alpha_i =0.01$. (a) The subluminal $u_\mathrm{FF}<c_8$
part of the flux-flow characteristics. Inset shows $I-V$
characteristics of individual junctions. Vertical lines show
positions of all possible Fiske steps. Appearance of various
superluminal resonances is indicated. (b) The $I-V$ in a wider
range. Dashed lines represent QP branches. Very strong
superluminal Fiske steps are seen. Panel (c) shows the $I-V$
characteristics of a mesa with nonuniform $I_{ci}$ and
$\alpha_i\simeq 0.05$, both increase by $5\%$ per IJJ from the top
to the bottom of the mesa. It is seen that variation of junction
resistances strongly suppresses high-frequency resonances.}
\end{figure}

Figure~\ref{fig:Fig5}(a) and (b) show simulated $I$-$V$ curves
(with subtracted top junction resistance) at $\Phi/\Phi_0=4.0$ for
the case of uniform damping $\alpha_i=0.01$, $(i=1,2,...8)$. Panel
(b) shows the total dc-voltage in a wide current range (current is
normalized by the total QP resistance of the mesa).
%QP resistances of eight active junctions are shown by dashed lines.
Basically all geometrical resonances with velocities from $c_8$ to
$c_1$ are seen, consistent with previous
simulations.\cite{FFlowSimul,FFlowMachida}

Panel (a) shows details of the subluminal flux-flow branch,
$u_\mathrm{FF}<c_8$. Here also a large variety of Fiske steps is
seen. The most frequent step separation is $V_{1,8}$ and
$2V_{1,8}$ (the grid spacing). The largest collective step occurs
at $8V_{7,8}$ close to the velocity matching condition,
Eq.~(\ref{VelMatch}). However, a finer step structure with
different voltages is developing in the top half of the flux-flow
branch at $V>2.5\,\mathrm{mV}$. To understand its origin, in the
inset we show individual voltages for all junctions. Vertical
lines indicate positions of all possible Fiske steps $V_{m,n}$.
The most prominent modes $(m,n)$ are indicated. As expected, in
the considered case of integer $\Phi/\Phi_0$ only odd $m=1,3,5,7$
resonances are excited. The observed even $2V_{1,8}$ step
separation is due to sequential switching between nearest odd-$m$
steps, as discussed in connection with Fig. \ref{fig:Fig3}(c).
However, not all steps in the subluminal regime are due to
out-of-phase $n=8$ resonances. Several superluminal resonances,
$n=3,4,5,6$, are also excited.

%******* FIGURE *******
%******* END FIGURE *******

Our simulations demonstrate that superluminal resonances $c_n>c_N$
can be excited by subluminal flux-flow $u_\mathrm{FF}<c_N$, which
is in agreement with experimental data reported above. Although
such possibility was not discussed before, it is probably not
surprising because qualitatively the same happens with
out-of-phase resonances. The necessary condition is given by
Eq.~(\ref{Mres}), which clearly allows excitation of superluminal
modes $n<N$ at $u_\mathrm{FF}<c_N$. For example, in the considered
case $N=8$, the first fastest $(1,1)$ mode can be excited at
$u_\mathrm{FF} \simeq c_8$, provided $\Phi/\Phi_0
> c_1/2c_8 \simeq 3$.

\subsection{Quality factors of different modes}
%**********************
\label{sec:qualityFactor} The quality factor $Q=\omega R C$ is an
important characteristic of the resonance. Here, $R$ is the
effective damping resistance and $C$ the capacitance of the
junction. High-quality geometrical resonances are crucial for
realization of coherent FFOs:\cite{TheoryFiske}
%******* ENUMERATE *******
\begin{enumerate}[i)]
%***
\item They amplify the emission power $P\propto E_\mathrm{ac}^2
\propto Q^2$. This is especially important for stacked FFOs since
there is no Lorentz contraction of the fluxon at the velocity
matching condition,\cite{Fluxon} which is used for amplification
of the electric field in a single junction FFO.\cite{Koshelets}
%***
\item They reduce the radiation linewidth of the FFO $\propto 1/Q$.
%***
\item  They may impose their order on the fluxon lattice even at
subluminal flux-flow velocities, as follows from numerical
simulations in Fig.~\ref{fig:Fig5}(a). In particular, in-phase
resonances may stabilize the rectangular fluxon lattice,
facilitating coherent amplification of radiation $P\propto Q^2
N^2$. The forced transformation of the fluxon lattice occurs when
the resonant standing wave electric field $E_\mathrm{ac}\propto
1/Q$ is larger than the fields of the fluxons. Therefore, high $Q$
is critical for such forced transformation.
\end{enumerate}
%******* END ENUMERATE *******
Although numerical simulations presented above describe well the
subluminal $u_\mathrm{FF}<c_8$ part of experimental flux-flow
branches [cf. Fig.~\ref{fig:Fig5}(a) and Fig.~\ref{fig:Fig3}],
there is a clear discrepancy in the superluminal regime. Numerical
simulations\cite{FFlowSimul,FFlowMachida} predict very strong
superluminal Fiske steps, which have even larger amplitudes than
the out-of-phase steps at $u_\mathrm{FF} \leq c_8$, as seen from
Fig.~\ref{fig:Fig5}(b). In those simulations larger $Q$ for higher
Fiske steps is simply the consequence of linear growth of
$Q\propto \omega$ for constant $R$ and $C$. To clarify this
discrepancy we consider the quality factors of different modes
$(m,n)$,
%*** EQUATION ***
\begin{equation}\label{Quality}
Q_{m,n} = \frac{\pi c_n m R(\omega)C}{L}.
\end{equation}
%*** END EQUATION ***
Here $L$ is the junction length. There are two counteracting
contributions. On the one hand, $Q\propto \omega_{m,n}$ and should
be larger for high speed and large $m$ modes. On the other hand,
the effective resistance $R(\omega)$ is decreasing with increasing
frequency. For single Josephson junctions it is very well
established that at high frequencies $R(\omega)$ is approaching
the radiative resistance of the circuitry $R_\mathrm{rad} \sim
50\,\Omega$\cite{Martinis} due to predominance of radiative losses
into outer-space.

As emphasized in Ref.~[\onlinecite{TheoryFiske}], in stacked
junctions $R(\omega)$ strongly depends on $n$. Out-of-phase modes
interfere destructively, cancelling radiative losses. Therefore,
$Q$ of those non-emitting modes is determined by the large QP
resistance $R_\mathrm{QP}$:
%*** EQUATION ***
\begin{equation}\label{QualityOP}
Q_{m,N} = \frac{\pi c_N m R_\mathrm{QP}C}{L}.
\end{equation}
%*** END EQUATION ***
Taking parameters for our mesa~1, $c_N\simeq 3.17\times
10^5\,\mathrm{m/s}$, $R_\mathrm{QP} \simeq 2\,\mathrm{k\Omega}$ at
$T=1.6\,\mathrm{K}$, intrinsic capacitance $C\simeq
70\,\mathrm{fF/\muup m^2}$, $L=2.7\,\mathrm{\muup m}$,
$L_y=1.4\,\mathrm{\muup m}$, and $m=2$ we obtain $Q_{2,8}\simeq
390$. This very large value is consistent with the sharp, large
amplitude $(2,8)$ steps seen in Fig.~\ref{fig:Fig3}(a). It is also
consistent with observation of out-of-phase Fiske steps in much
longer junctions $L=50\,\mathrm{\muup m}$.\cite{Fiske} The same is
true for other destructively interfering, non-radiative modes
%*** EQUATION ***
\begin{equation}\label{Mnonrad}
n^*=N-2i,\qquad i=0,1,2,\ldots, \mathrm{int}\left[(N-1)/2\right].
\end{equation}
%*** END EQUATION ***

For the in-phase mode electric fields from all junctions interfere
constructively resulting in large coherent emission. In this case
$R$ is determined by radiative losses,
%*** EQUATION ***
\begin{equation}\label{QualityIP}
Q_{m,1} = \frac{\pi c_1 m R_\mathrm{rad} C}{L}.
\end{equation}
%*** END EQUATION ***
Using parameters of mesa 1, $c_1\simeq 1.8\times
10^6\,\mathrm{m/s}$, and assuming
$R_\mathrm{rad}=50\,\mathrm{\Omega}$ we obtain $Q_{2,1} \simeq
46$. Thus, radiative geometric resonances should have
substantially smaller $Q$ than the non-radiative
ones.\cite{TheoryFiske}

For mesa~1 with $N=8$ all even-$n$ modes are non-radiative,
Eq.~(\ref{Mnonrad}). Fiske steps due to such resonances should
have the highest $Q$ and amplitude. This is consistent with our
observation that even-$n$ modes correspond to the most prominent
Fiske steps, see Fig.~\ref{fig:Fig3}. Similarly, more smeared,
high-voltage Fiske steps at larger fields could indicate enhanced
emission and radiative losses from the radiative superluminal
resonances. The asymmetry between even and odd $n$ Fiske steps is
an indirect evidence for significant coherent radiation emission
from the stack.\cite{TheoryFiske}

\subsection{Suppression of superluminal resonances by the spread in junction resistances.}
%**********************
\label{sec:ResSpread} Collective, high-voltage Fiske steps can
also be suppressed by another trivial reason: due to the spread in
junction resistances $\Delta R$. Indeed, if junctions in the stack
have slightly different QP resistances $R_i$, the Fiske step
$V_{m,n}$ would occur at different currents $I_i=V_{m,n}/R_i$.
Therefore, the collective Fiske step could be observed only if the
amplitude of the steps $\Delta I > (V_{m,n}/R_i^2) \Delta R$.
Apparently, this condition is more difficult to satisfy for high
voltage and low $Q$ resonances.

To demonstrate this, in Fig. \ref{fig:Fig5}(c) we present
numerical simulations for nonuniform damping $\alpha_i \simeq
0.05$, increasing by $5\%$ per IJJ from the top to the bottom of
the mesa and with the corresponding spread in the QP resistances
$R_i \propto 1/\alpha_i$. The rest of the parameters are similar
to that in panels (a) and (b). From comparison with panel (b) it
is clear that the amplitude of low frequency out-of-phase $c_N$
steps $IR_n \simeq 10$ mV is unchanged, but all high frequency
superluminal resonances are strongly reduced. The fastest in-phase
resonances are practically nonexisting. We emphasize that in the
considered case this is not due to radiative losses, but is caused
by the trivial reason that junctions can not be synchronized
because it is impossible to keep them at the same voltage for a
given current. Note that the spread in $I_c$ is not preventing the
synchronization and is, therefore, much less detrimental for
resonances, as seen from Fig. \ref{fig:Fig5}(b). In practice, both
low $Q$ due to radiative losses\cite{TheoryFiske} and enhanced
sensitivity to resistance spread may explain the abundance of the
highest speed Fiske steps in experiment.

\section*{Conclusions}
%**********************
\label{sec:conclusions} In conclusion, careful alignment of
magnetic field allowed observation of a large variety of Fiske
steps in small Bi-2212 mesas. Small number of IJJs in the mesa
allowed accurate identification of different resonant modes.
Different resonance modes, including superluminal with velocities
larger than the lowest velocity of electromagnetic waves $c_N$,
were observed. It was shown both experimentally and theoretically
that superluminal geometrical resonances can be excited in the
subluminal flux-flow state. Superluminal flux-flow state with the
maximum flux-flow velocity up to the Swihart velocity $c_0 \simeq
1.4 c_N$ was also reported. The most prominent observed Fiske
steps correspond to non-emitting resonance modes $n^*$, Eq.
(\ref{Mnonrad}). The corresponding asymmetry between even and
odd-$n$ resonances can be viewed as indirect evidence for
significant coherent emission from intrinsic Josephson
junctions.\cite{TheoryFiske}
%We have argued that high-quality superluminal resonances are crucial for realization of a coherent flux-flow oscillator in the THz frequency range.
\vspace{0.001mm}%Fixing spacing problem of acknowledgments

%******* ACKNOWLEDGEMENTS *******
\begin{acknowledgments}

We are grateful to A.~Tkalecz for assistance with sample
fabrication. The work was supported by the K.~\&~A. Wallenberg
foundation, the Swedish Research Council and the SU-Core Facility
in Nanotechnology.
\end{acknowledgments}

%******* REFERENCES *******

\end {document}